\documentstyle[epsfig, twocolumn]{venice97}

\newcommand{\muas}{\hbox{$\mu${\rm as}}}

\begin{document}

%% Do not remove the following six lines:
\setlength{\parindent}{0pt}
\setlength{\parskip}{ 10pt plus 1pt minus 1pt}
\setlength{\hoffset}{-1.5truecm}
\setlength{\textwidth}{ 17.1truecm }
\setlength{\columnsep}{1truecm }
\setlength{\columnseprule}{0pt}
\setlength{\headheight}{12pt}
\setlength{\headsep}{20pt}
\pagestyle{veniceheadings}

%% Title - should be in capitals:
\title{GAIA AND THE HUNT FOR EXTRA-SOLAR PLANETS}
\author{{\bf M.G. Lattanzi,$^1$ A. Spagna,$^1$ A. Sozzetti,$^2$ 
    S. Casertano,$^3$}\thanks{Affiliated to the Space Sciences Dept., ESA}
    \vspace{2mm}\\
    $^1$Osservatorio Astronomico di Torino, Pino Torinese TO, I-10025, Italy\\
    $^2$Dipartimento di Fisica Generale, Universit\'a di Torino, I-10100, Italy\\
    $^3$Space Telescope Science Institute, Baltimore MD, 21218, USA}

\maketitle

\begin{abstract}

We present the results of realistic end-to-end simulations of
observations of nearby stars with the proposed global astrometry mission
GAIA, recently recommended within the context of ESA's Horizon 2000 Plus
long-term scientific program.  We show that under realistic, if
challenging, assumptions, GAIA will be capable of surveying the solar
neighborhood within 100-200 pc for the astrometric signatures of planets
around stars down to $ V = 16 $ mag. 

The wealth of results on the frequency and properties of massive planets 
from GAIA observations will provide a formidable testing ground on
which to confront the most sophisticated theories on planetary
formation and evolution. 

Finally, we suggest the possibility of more sophisticated probabilistic
detection techniques which may be able to detect the presence 
of Earth-like planets around stars within 20 pc. 
\vspace {5pt} \\

Key~words: astrometry; stars:planetary systems; GAIA.

\end{abstract}

\section{INTRODUCTION}

After many years of work, radial velocity searches have finally 
succeeded in finding a number of Jupiter-like planets orbiting nearby 
stars.  Currently, there exist about 10 candidate planets, all found by 
the radial velocity technique around stars of solar type or later
(see e.g.\ \cite{mayor95}, \cite{marcy96}, \cite{butler96}).

However, most candidate planetary systems identified thus far seem to
defy the prior expectations of well-established theoretical explanations
of the formation and evolution of planets.  Giant planets appear to have
either very short periods (\cite{mayor95}), placing them well outside the 
freezing zone where their formation was expected to take place, or have
high eccentricity (\cite{cochran97}), again in contrast with the
planetesimal accretion model. 

Astrometric techniques are complementary to the radial velocity
detection currently employed, in that, for given mass, the sensitivity
of astrometry and radial velocity techniques respectively increases and
decreases with period.  Thus, astrometric searches will find
preferentially planets at several AU from the central star, with periods
of several years, as opposed to radial velocity searches which favor
planets very close to the central star. 

A GAIA-like satellite, capable of extremely high precision astrometric
measurements, will revolutionize our knowledge of planetary systems and
provide invaluable statistical information for theoretical formation
models.  With the baseline properties currently envisioned
(\cite{lindegren96}), GAIA will find more than 50 per~cent of all Jupiter-like
planets orbiting stars within 100 pc with periods between 1 and 15 years,
averaged over eccentricity, inclination, and orbital phase, as well as
lower-mass planets around more nearby stars.  
The orbital elements of many detected systems can be evaluated,
including mass and orbital radius, unlike radial velocity
detection which leaves significant uncertainty in the two quantities
because of the inclination uncertainty. If solar-like planetary systems
are reasonably common, we can expect hundreds of thousands of detections,
providing ample material to test both theoretical predictions and
detailed models.  Even detection of Earth-like planets could become
viable in a statistical sense, under appropriate circumstances. 

\section{DATA SIMULATION} 

The simulation code is adapted from that used by \cite*{galligani89} for
the assessment of the astrometric accuracy of the sphere reconstruction
in the Hipparcos mission.  We generate catalogs of single stars randomly
distributed on the sky; each run produces a sphere of $N$ stars,
with the same value of parallax, total proper motion, magnitude and color.
%
%For the
%moment, parallaxes and proper motions, as well as magnitudes and colors,
%are drawn from simple distributions which do not represent any
%particular Galaxy model; in particular, in each run the 100 stars
%simulated have the {\it same} parallax, proper motions, magnitudes, and
%colors. 

The simulations presented here are carried out within the great circle
approximation (\cite{lindegren89}).   
We neglect all difficulties related to the
reconstruction and calibration of individual great circles, difficulties
that would be more properly addressed within the context of a global
model of the Galaxy, and consider as the {\it basic observable} the abscissa $
\psi $ along the instantaneous great circle measured in each observation
of each star.  The accuracy expected for individual measurements of $
\psi $ has been discussed in \cite*{casertano96}, whose basic
assumptions we adopt here.  The measurement error expected for
individual observations depends on the instrumental parameters, here
taken as in \cite*{lindegren96}, on the magnitude of the star and on the
quality of the metrology control; the results of \cite*{casertano96} are
reported in Table~\ref{errors} for reference. 

The new element added here is that the instantaneous ``true'' position
of each star includes the gravitational perturbation (Keplerian motion)
induced by a single, non-luminous planetary mass orbiting the star. 

\begin{table}
\caption{Photon and total error for a single observation (10 
elementary exposures)}
\label{errors}
\catcode`\!\active \def!{\hbox{\phantom{0}}}
\begin{tabular}{@{}cccc}
$ V $ mag & Photon error
&\multicolumn{2}{c}{Total error ({\muas})}\\
%\cline{3-4}
 & ({\muas}) & ($\sigma_{\rm b}$ = 200 pm) & ($\sigma_{\rm b}$ = 20 pm) \\
%\noalign{\smallskip\hrule\smallskip}
 !0   &   !!0.032   &   !!5.3   &   !!0.53 \\
 !1   &   !!0.050   &   !!5.3   &   !!0.53 \\
 !2   &   !!0.080   &   !!5.3   &   !!0.54 \\
 !3   &   !!0.126   &   !!5.3   &   !!0.55 \\
 !4   &   !!0.20!   &   !!5.3   &   !!0.57 \\
 !5   &   !!0.32!   &   !!5.3   &   !!0.62 \\
 !6   &   !!0.50!   &   !!5.3   &   !!0.73 \\
 !7   &   !!0.80!   &   !!5.4   &   !!0.96 \\
 !8   &   !!1.3!!   &   !!5.5   &   !!1.4! \\
 !9   &   !!2.0!!   &   !!5.7   &   !!2.1! \\
 10   &   !!3.2!!   &   !!6.2   &   !!3.2! \\
 11   &   !!5.0!!   &   !!7.3   &   !!5.1! \\
 12   &   !!8.0!!   &   !!9.6   &   !!8.0! \\
 13   &   !12.6!!   &   !13.7   &   !12.6! \\
 14   &   !20.0!!   &   !20.7   &   !20.0! \\
 15   &   !31.7!!   &   !32.2   &   !31.8! \\
 16   &   !50.3!!   &   !50.6   &   !50.3! \\
 17   &   !79.7!!   &   !79.9   &   !79.7! \\
 18   &   126.4!!   &   126.5   &   126.4! \\
 19   &   200.3!!   &   200.4   &   200.3! \\
 20   &   317.5!!   &   317.5   &   317.5! \\
\end{tabular}
\end{table}

\section {DETECTION OF JUPITER-LIKE PLANETS}

\subsection {Star luminosity and measurement error}

We consider first the case of planets with mass comparable to that of
Jupiter.  Such planets produce a relatively large astrometric perturbation;
at 10~pc the reflex motion of the Sun due to Jupiter's motion would be 
500~{\muas} with a period of 11.8 years.  
%On the other hand, Jupiter's relatively long period
%(11.8~yr) would make it somewhat harder to detect in a short mission,
%since the period is incompletely sampled.  
We will parametrize the
detection probability by the two major contributors, 
the period $P$ and the signature 
\begin{equation}
\alpha=\frac{M_{\rm p}}{M_{\rm s}} \; \frac{a_{\rm p}}{D} 
\end{equation}
where $M_{\rm p}$, $M_{\rm s}$ are the masses of the planet and star
respectively, $a_{\rm p}$ the  semimajor axis of the planetary system,
and $D$ its distance from us.   If $a_{\rm p}$ is in AU, and 
$D$ in parsec, then $\alpha$ is expressed in arcseconds.
%, and average the probability over the
%other orbital parameters.  
We generally assume a single-observation
measurement error $\sigma_\psi= 10$ {\muas},   
appropriate to a star brighter than $ V
\sim 12 $~mag (see Table \ref{errors}), corresponding to the Sun at
200~pc.  
%However, we demonstrate explicitly that the detection
%probability depends exclusively on the ``signal-to-noise'' ratio $ \alpha
%/ \sigma_\psi $, and thus rescaling to different measurement errors is
%straightforward. 

We also carried out tests with different measurement errors, 
and demonstrated explicitly that the detection
probability depends exclusively on the ``signal-to-noise'' ratio 
\[
 S/N = \alpha / \sigma_\psi 
\]
so that rescaling to different measurement errors is straightforward.

\subsection{Detection method} 

Our detection method for Jupiter-like planets is a classical
application of the $ \chi^2 $ test.  After assigning each star in the
simulation a secondary component, we solve for the five astrometric
parameters for that star as if it had no companion.  We then apply a
standard $\chi^2$ test (with the confidence level set to 95~per~cent) to
the residuals $\psi - \psi_r$, where the $\psi$ are the actual
measurements, and the $\psi_r$ the great circle abscissae recomputed on
the basis of the single-star fit. If the test is failed, that is, the
residuals are significant at the 95~per~cent level, the planet is
considered ``detected''.  Note that this method measures only deviations
from the single-star model and makes no assumptions on the nature of the
residuals, nor does it give an indication of whether the planet's
parameters can be computed. 

The simulated data for this case include a sample of 160\,000 stars
uniformly distributed over the sky.  Each star is assumed to have a
planet inducing a reflex astrometric motion of amplitude $ \alpha $
ranging from 5 to 100 {\muas}, and period $ P $ between 0.5 and 20
years. The remaining orbital elements are distributed randomly in the
ranges: $0^\circ\le i\le 90^\circ$, $0\le e\le 0.3$, $0\le \Omega\le 2\pi$, 
$0\le \omega\le 2\pi$, $0\le T\le P$. 
%and the detection probability is averaged over them. 

This means that the detection probability discussed in this
study must be considered as averaged on both the mission parameters (such
as number of observations vs ecliptic latitude) and on the orbital elements
(e.g.\ inclination and eccentricity).
Detailed investigations of
these detection methods and their dependence on those parameters are 
in progress and will be presented elsewhere. 

Finally, since the ``detection'' of a planet is indicated by a $ \chi^2 $
deviation significant at the 95 per~cent level, we would expect a
5~per~cent incidence of false detections.  As a check, we repeated the
simulation without planets, and we did in fact find false detections
consistent with the expected 5~per~cent. 

\subsection {Results}

\begin{figure}[t!]
\centerline {\epsfig{file=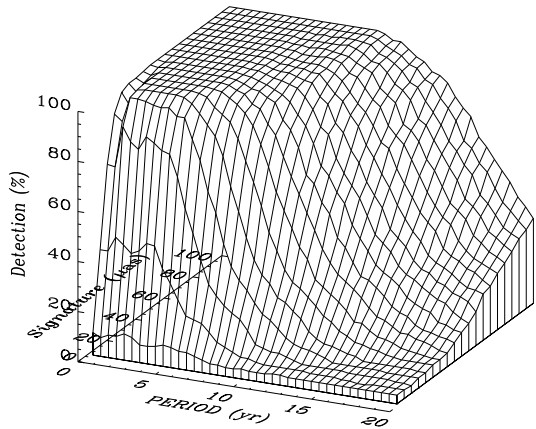,width=9.0cm}}
\caption{\em 
   Planet detection probability as function of the astrometric signature
   $\alpha$ and of the orbital period $P$, for $\sigma_\psi$ = 10 {\muas}.
   The percentage of detection of each point is based on 200 random planetary
   systems uniformly distributed on the sky.}
\label{tred}
\end{figure}

The fraction of planets detected---as measured by the failure of the $
\chi^2 $ test for the single-star hypothesis---is given in Figure
\ref{tred} as a function of orbital period and $ \alpha $, for an
assumed measurement error of $ 10 \muas $.  We note that at relatively
low S/N ratios the detection probability is dominated by sampling of the
orbital period, while at higher S/N values orbital sampling is less
critical and long period planets (up to about twice the mission
duration) are detectable. 
For instance, the detection probability reaches
about cent per cent when S/N $\rightarrow$ 10.

Figure~\ref{tred} also shows that for
S/N$\rightarrow$ 1 the $\chi^2$ test quickly loses its sensitivity. 
The shallow dip in the detection probability at $ P \sim 1 $ year is the 
result of the coupling between orbital and parallactic motion.

As stated above, the results shown in Figure \ref{tred} can be scaled
easily to other measurement accuracies, whether due to different
assumptions on the properties of the mission or to different stellar
magnitudes; for example, for measurement accuracy $ \sigma_\psi = 1
\muas $, the detection probability is exactly the same as shown in the
Figure \ref{tred}, but for an amplitude ten times smaller---thus maintaining the
same S/N ratio. 

\subsection {Parameters for 50~per~cent detection}

Another way to look at these results is to 
determine the amplitude of the perturbation needed for a 
certain probability of detection, as a function of the planet's 
period, and compare this relation with the Kepler's third law, $\alpha\propto
P^{2/3}$, which depends explicitly on the physical parameters of the
planetary system.

\begin{figure}[t!]
\centerline {\epsfig {file=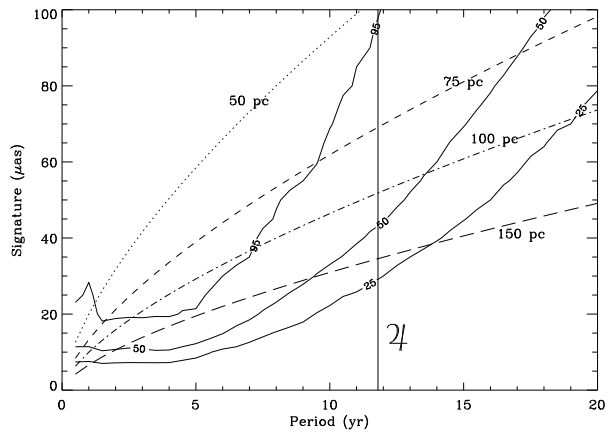,width=8.5cm}}
\caption{\em Iso-probability contours ({\it solid lines}) for 25\%, 50\% and 95\% 
of detection probability, compared with Kepler's third laws 
({\it dotted/dashed lines}) for  systems with Jupiter-Sun masses
at $D=50$, 75, 100 and 150 pc.
%
%The empirical signature-period relation for Jupiter-mass
%planets found in this study (solid line). The dashed lines refer to the
%physical relations derived from Kepler's third law for different
%distances of the systems. 
Jupiter-like planets ($P=11.8$ yr) appear detectable, with
probability $\ge$ 50\%, up to a distance of 100 pc (vertical line).} 
\label{sp50}
\end{figure}

 The empirical relations derived from Figure~\ref{tred} are shown 
in Figure~\ref{sp50} for three levels of detection efficiency (25, 50,
95 per~cent).  
Orbital periods shorter than 5 years are 
well-matched to the mission length and sampling law, and 
therefore the detection probability is nearly independent of orbital 
period, with a detection probability of 50~per~cent  when $ S/N 
\sim 1$.
% {\em i.e.}, when the perturbation amplitude $ \alpha $ is about 
%twice the single-measurement error $ \sigma_\psi $ (\cite{casertano95} 
%assumed detection if $ S/N \sim 3 $ in one-year normal points, which 
%corresponds closely to the condition found here).  
On the other
hand, if the period exceeds the mission lifetime, the 
probability of planet detection drops significantly, and a much higher 
signal is required for the planet's signature to be detected.  This is 
in qualitative agreement with the results of \cite*{babcock94},  
% for the mission POINTS, although there are some minor differences in the 
%quantitative results obtained, especially in the dependence on the 
%planetary period which appears steeper in the \cite*{babcock94} results.
who studied the detection {\em and} convergence probability of a complete 
orbital model
for simulated planetary systems as observed by the mission POINTS.
\cite*{babcock94} did find a slightly larger sensitivity on planet period,
manifested in an earlier turn-up and steeper slope at long periods of the
50 per~cent probability curve; this most likely depends on the fact that the
determination of reliable orbital elements is more challenging than detection
only.

A planet with exactly the same characteristics as Jupiter ($P=11.8$~yr) can be 
detected with probability greater than 50~per~cent to a distance of 
100~pc,  
%The reason is the Jupiter's orbital period of 11.8~yr (vertical 
%dotted line in Figure~\ref{sp50}) is long compared with the mission 
%lifetime, planned at 5 years.  On the other hand, 
while planets with mass similar
to Jupiter's, but shorter orbital periods, can be detected much further
away, to distances exceeding 150~pc for periods between 2 and 9~yr.  
Depending on the orbital period, the number of candidate stars for 
detection of Jupiter-like planets may well be several hundred thousand 
(see also \cite{casertano95}).

\section {DETECTABILITY OF KNOWN CANDIDATE PLANETS}

As a further test of GAIA's capabilities, we consider explicitly three
candidate planets discovered by the radial velocity technique, 47 Uma,
70 Vir, and 51 Peg.  The parent stars of all three systems are close by
($d\le$ 20 pc) and very similar to our Sun. According to the
spectroscopic measurements summarized in \cite*{perryman96}, the planets 
orbiting these
stars have minimum masses of $\sim 2.46 $, $\sim 6.50 $, and 
$\sim 0.5 M_{\rm J} $, and orbital periods of about 3 years, 4 months,
and 4 days, respectively.  This translates in the following minimum
astrometric signatures: 

\begin{center}
\begin{tabular}{@{}rcl}
$\alpha_{\rm 47 UMa}$ & $\ge$ & 362 \muas  \\
$\alpha_{\rm 70 Vir}$ & $\ge$ &168 \muas  \\
$\alpha_{\rm 51 Peg}$ & $\ge$ &1.66 \muas  \\
\end{tabular}
\end{center}

The astrometric detection of 51 Peg by a GAIA-like mission is extremely
difficult, because of the small astrometric signature (short period
implies small separation, thus small reflex motion) and of the mismatch
between the orbital period and the frequency of GAIA observations.
On the other hand, the detection of the signatures induced on 47 UMa 
and 70 Vir should be a much easier task for a GAIA-like satellite. 

In the context of a new class of simulations, we have generated 100
planetary systems on the celestial sphere, respectively identical to 47
UMa and 70 Vir, assuming the stars to be of mass M = M$_\odot$, and
assuming perfectly circular orbits. The inclination of the orbital
planes, undetermined parameter in the case of radial velocity
measurements, was initially chosen to be $i=45^\circ$. 

The $\chi^2$ test indicated a detection probability of essentially 
100~per~cent for these planets for a single-observation error of 
$\sigma_\psi=10 \muas $ (a conservative assumption for these 
relatively bright stars).  We were also able to recover accurately 
all orbital elements and to reconstruct the apparent path of the stars 
on the plane of the sky.  The details of these simulations will be 
published elsewhere.

\section{DETECTION OF EARTH-LIKE PLANETS}

Even for very nearby stars, within 10~pc or so, detection of Earth-like 
planets will be extremely challenging.  The astrometric signature of the 
Earth on the position of the Sun corresponds to about $ 0.3$~\muas\  at 
10~pc, beyond the capabilities currently projected for GAIA.

The question we address here is whether it may be possible to establish 
{\it statistically} the presence of Earth-like planets, even though they 
cannot be detected directly on an individual basis.  Specifically, we 
consider the possibility that the statistical properties of the 
residuals for a few hundred stars might bear a weak signature of the 
presence of Earth-like planets, and that, by combining these data, the 
evidence for the presence of such planets might be uncovered.

To this end, we use the least-squares technique in a non-conventional
way. Contrary to what we have done before, we {\it assume} the presence
of a planet, and fit the observations with a model which includes the
semi-major axis, $a$, of the stellar orbital motion along with the five
astrometric parameters. We repeat this process for different values of
the period $ P $. Convergence of the fit, and not the significance of
the values derived for $a$, is taken as the indicator of a positive
detection.\footnote{As expected, because of the low S/N ratios, the
error of the estimated values of $a$ is usually large and not very
significant.} 
Observations for Earth-Sun systems were simulated with an error
$\sigma_\psi=1$~\muas\, as indicated in Table~\ref{errors} for nearby
solar stars, in the case of a metrology accuracy of 20~pm.
As for the case of the $\chi^2$ test, the same simulations
were repeated without astrometric signatures in order to assess the
probability of false detections. 

\begin{figure*}[t!]
\centerline {\epsfig{file=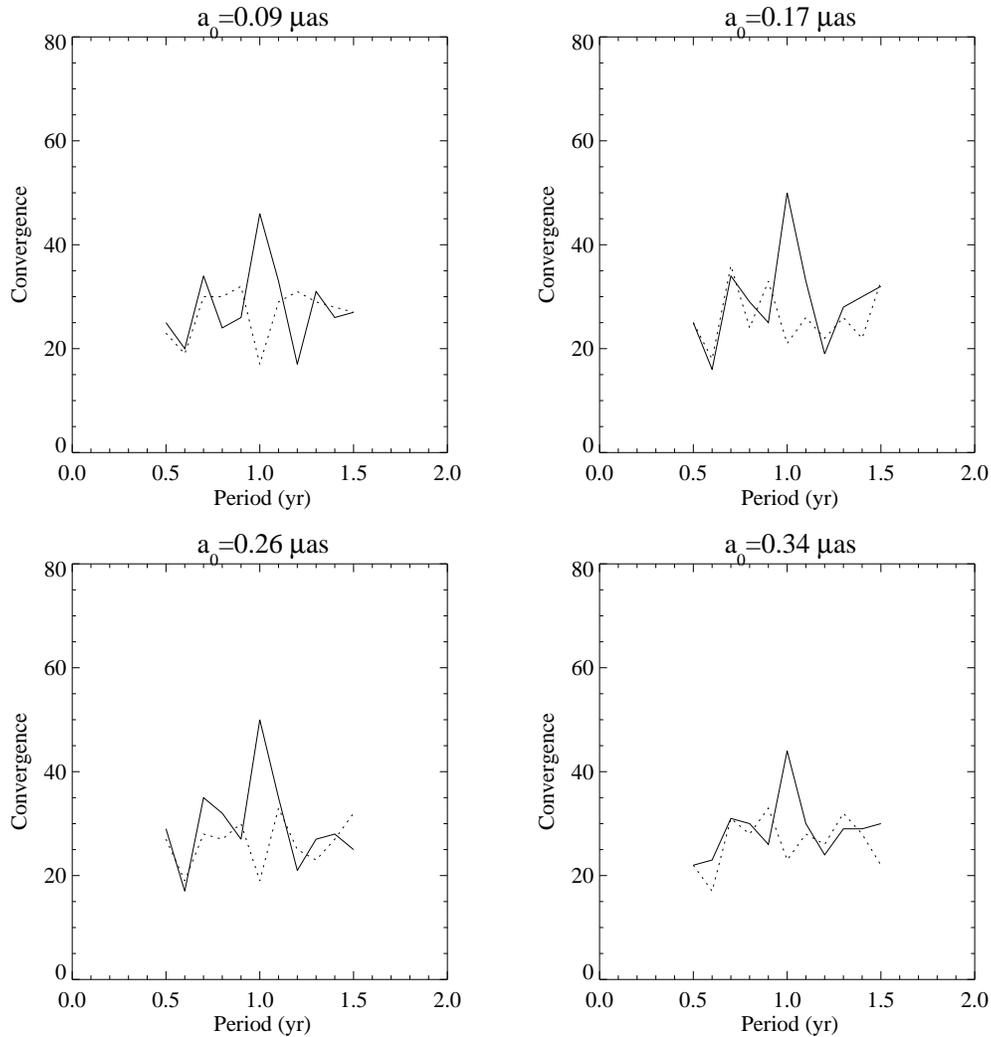,width=14cm}}
%\centerline {\epsfig{file=detect_earth_20PC.PS,width=14cm}}
\caption{\em Convergence probabilities for Sun-Earth planetary systems
(solid curve) at 20 pc as function of the initial guesses on the
one-year period. The dashed curve shows the number of false detection.
Panels refer to different starting values of the orbital radius $a_0$;
the correct value is $a=0.17$~\muas.
 The 22 data points used to plot both graphs 
%represents an average over 4 different simulations 
are based on 100 stars each.} 
\label{earth20}
\end{figure*}

We find that the probability of convergence does in fact depend on the
presence of the planet.  As shown in Figure \ref{earth20} for systems
at 20 parsec, 
fitting a system without a planet or with the incorrect period results in a
smaller convergence probability than if the planet is present and the
period correct.  For an initial value $ a_0 = 0.34 \muas $, the fit
converges only 25~per~cent of the time if there is no planet or if the
planet has a period different by as little as 0.05~yr from the period
used in the fit.  On the other hand, if a planet with the correct period
is present, the fit converges about 45~per~cent of the time.   This
indicates that a statistical signature of the presence of an Earth-like
planet is indeed present, albeit weak.  Of course, this cannot lead to
the actual detection of individual planets.  However, given a sufficient
number of candidates---there are about 400 eligible stars within
20~pc---this method might be used to detect whether, and how often,
Earth-like planets may be present, even though the S/N of the signature
of individual planets may be insufficient for a more formal detection. 
Of course, there will be little or no information on the detailed
orbital parameters of such planets, but the interest, scientific and
not, of the detection of Earth-like planets is such that even
probabilistic detections are desirable. 

The current analysis is still too simplistic to assess whether this
method, or variants thereof, can be successful.  We need to investigate
further the effect of the other parameters---initial phase, inclination,
eccentricity, etc.---which for now have been assumed ``known''; initial
phase may have an especially large effect on the convergence of the fit.
Similarly, a method based on the statistical analysis of low S/N cases
will only be successful if the error properties of the measurements are
extremely well-known, and if the possible presence of other planets
does not affect the convergence of the fit. On the other hand, the
ability to ``detect'' planets with $ S/N < 1 $  is very tantalizing, and
well worth of further study. 

\section{SUMMARY AND CONCLUSIONS}

Detailed simulations of observation of star-planet systems within the 
baseline framework of the proposed GAIA mission indicate the probable 
discovery of a very large number of massive, Jupiter-like planets around 
stars as far away as 100-200~pc, thus enabling a qualitative jump in the 
statistical study of planetary systems and new understanding   
of their formation.  Specific simulations of known candidate planetary 
systems, discovered by the radial velocity technique, indicate 
that such systems will be easy to discover and their orbital parameters 
will be determined accurately by GAIA, with the exception of very 
short-period systems such as 51 Pegasi.  

\section*{ACKNOWLEDGMENTS}
We wish to thank M.A.C. Perryman for his continuing support of this
investigation.

\end{document}